\documentclass[letter,twocolumn]{jpsj3}
\usepackage{graphicx} 
\usepackage{tabularx}
\usepackage{dcolumn} 
\usepackage{color}
\usepackage{bm}
\usepackage{amsmath}
\usepackage{amssymb}
\usepackage{multirow}
\newcommand{\bi}[1]{\ensuremath{\boldsymbol{#1}}} 
\newcommand{\nn}{\nonumber}
\title{ 
Coherence Effects of Caroli-de Gennes-Matricon Modes in Nodal Topological Superconductors
}

\author{Yasumasa Tsutsumi\thanks{tsutsumi@vortex.c.u-tokyo.ac.jp} and Yusuke Kato} 
\inst{Department of Basic Science, The University of Tokyo, 
Meguro, Tokyo 153-8902, Japan} 
 
\abst{
Coherence effects by the impurity scattering of Caroli--de Gennes--Matricon (CdGM) modes in a vortex for nodal topological superconductors have been studied.
The coherence effects reflect a topological number defined on a particular momentum space avoiding the superconducting gap nodes.
First, we analytically derived the eigenvalue and eigenfunction of the CdGM modes, including the zero-energy modes, in a nodal topological superconducting state without impurities, where we focused on a possible superconducting state of UPt$_3$ as an example.
Then, we studied impurity effects on the CdGM modes by introducing the impurity self-energy, which are dominated by the coherence factor depending on the eigenfunction of the CdGM modes.
For the zero-energy CdGM modes, the coherence factor vanishes in a certain momentum range, which is guaranteed by topological invariance characterized by the one-dimensional winding number.}
  
\begin{document} 
\maketitle 


Topological materials, such as quantum Hall systems~\cite{prange:book}, topological insulators~\cite{hasan:2010,qi:2011}, and topological superconductors/superfluids~\cite{qi:2011,mizushima:2015,mizushima:2016}, have been attracting wide attention, and investigation of their topological phases has been developing beyond the field of condensed matter physics~\cite{volovik:book}.
The topological phase is characterized by topological numbers defined in gapped states, where topological invariance guarantees the presence of gapless edge modes.
Also for nodal superconductors/superfluids, the topological numbers can be defined on a particular momentum space avoiding gap nodes~\cite{beri:2010,sato:2010}.
However, when there is momentum transfer due to the presence of impurities, the validity of the procedure of taking a particular momentum space is not clear.

Zero-energy vortex modes in a topological phase are also characterized by topological numbers defined on a base space $({\bm k},\phi)$, which consists of the momentum and a parameter $\phi$ on a circle surrounding the vortex~\cite{teo:2010}.
In superconductors/superfluids, the zero-energy vortex modes are derived from the Caroli--de Gennes--Matricon (CdGM) modes~\cite{caroli:1964}.
In particular, on superconductors, the robustness of the CdGM modes, namely their relaxation time, can be observed by flux flow conductivity measurements~\cite{kopnin:book}.
Therefore, impurity effects on the CdGM modes are promising for providing a criterion for the availability of the topological classification of nodal superconductors.

The heavy fermion superconductor UPt$_3$~\cite{stewart:1984,joynt:2002} is a highly probable nodal topological superconductor.
As a result of numerous experimental and theoretical studies over three decades, the possible gap functions in UPt$_3$ have been narrowed to the $E_{1u}$ planar state~\cite{machida:2012,tsutsumi:2012b}, $E_{1u}$ chiral state~\cite{izawa:2014}, and $E_{2u}$ chiral state~\cite{choi:1991}.
All of the pairing states in the low-temperature and low-field phase, i.e., the B-phase, show topologically protected gapless edge modes~\cite{tsutsumi:2013,goswami:2015}.
In this paper, we mainly focus on the $E_{1u}$ planar state described by the $d$-vector ${\bm d}({\bm k})\propto({\bm x}k_y+{\bm y}k_x)(5k_z^2-k^2)$ as an example of the nodal topological superconducting state.
The gap function has point nodes at the north and south poles on the Fermi surface and two horizontal line nodes at $k_z=\pm k_{\rm F}/\sqrt{5}$, where $k_{\rm F}$ is the Fermi wave number.

Our aim is to clarify the significance of the topological classification of nodal superconductors by studying impurity effects on the CdGM modes.
In a vortex along the $z$-direction, the $E_{1u}$ planar state has many zero-energy CdGM modes with momentum $|k_z|<k_{\rm F}$ between the point nodes.
In this paper, we have demonstrated that the zero-energy modes show characteristic impurity effects owing to the coherence factor vanishing in a certain momentum range.
The coherence factor of the zero-energy modes reflects a topological number, which does not change unless crossing the horizontal line nodes.
Thus, the topological classification of nodal superconductors is effective even when momentum transfers are caused by impurity scattering.


First, we derive CdGM modes without impurities from the Bogoliubov--de Gennes (BdG) equation.
The BdG equation for an inhomogeneous order parameter is described by~\cite{kawakami:2011}
\begin{align}
\int d{\bm r}_2
\begin{pmatrix}
\hat{\epsilon }(\bi{r}_1,\bi{r}_2) & \hat{\Delta }(\bi{r}_1,\bi{r}_2) \\
\hat{\Delta }^{\dagger }(\bi{r}_2,\bi{r}_1) & -\hat{\epsilon }^{\rm T}(\bi{r}_2,\bi{r}_1)
\end{pmatrix}
\vec{u}_{\nu }(\bi{r}_2)
=E_{\nu }\vec{u}_{\nu }(\bi{r}_1).
\label{eq:BdG}
\end{align}
The normal-state Hamiltonian omitting the vector potential, which can be neglected when considering the CdGM modes in low fields with a large Ginzburg--Landau parameter $\kappa\gg 1$~\cite{caroli:1964}, is given by
\begin{multline}
\hat{\epsilon }(\bi{r}_1,\bi{r}_2)\\
=\delta(\bi{r}_1-\bi{r}_2)\frac{\hbar^2}{2m}\left[-\partial_{\rho }^2-\frac{1}{\rho }\partial_{\rho }-\frac{1}{\rho^2}\partial_{\phi}^2-\partial_z^2-k_{\rm F}^2\right]\hat{\sigma }_0,
\end{multline}
where $m$ is the particle mass, $\hat{\sigma }_0$ is the $2\times 2$ unit matrix, and $(\partial_{\rho },\partial_{\phi },\partial_z)$ are differential operators in cylindrical coordinates.
The pair potential is
\begin{align}
\hat{\Delta}(\bi{r}_1,\bi{r}_2)=\int\frac{d{\bm k}}{(2\pi)^3}\hat{\Delta }({\bm r},{\bm k})e^{i{\bm k}\cdot{\bm r}'},
\end{align}
with ${\bm r}=({\bm r}_1+{\bm r}_2)/2$ and ${\bm r}'={\bm r}_1-{\bm r}_2$,
and the wave function is
\begin{align}
\vec{u}_{\nu }({\bm r})=
\begin{pmatrix}
u_{\nu }^{\uparrow }({\bm r}) \\
u_{\nu }^{\downarrow }({\bm r}) \\
v_{\nu }^{\uparrow }({\bm r}) \\
v_{\nu }^{\downarrow }({\bm r})
\end{pmatrix}.
\label{eq:BdGwave}
\end{align}

\begin{figure}
\begin{center}
\includegraphics[width=8.5cm]{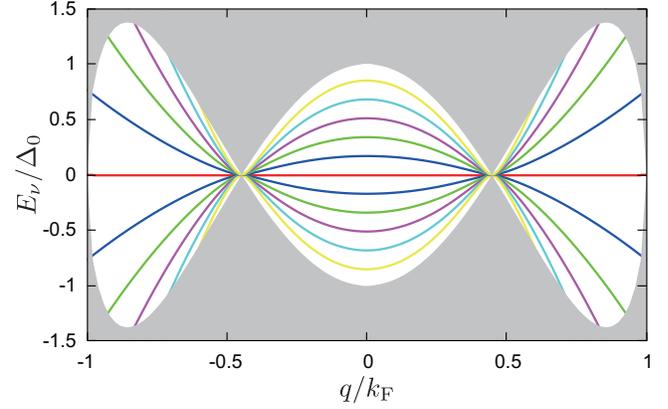}
\end{center}
\caption{\label{fig1}(Color online)
Eigenvalue of the CdGM modes $E_{\nu }=-l\omega_q$ for $|l|\le 5$ within the superconducting gap $\Delta_0|(5\cos^2\alpha -1)\sin\alpha|$, where $q/k_{\rm F}=\cos\alpha$.
}
\end{figure}

For the $E_{1u}$ planar state~\cite{machida:2012,tsutsumi:2012b}, the gap function is described by
\begin{align}
\hat{\Delta }({\bm r},{\bm k})\equiv&i{\bm d}({\bm r},{\bm k})\cdot\hat{\bm \sigma }\hat{\sigma }_y\nn\\
=&i\Delta({\bm r})({\bm x}k_y+{\bm y}k_x)(5k_z^2-k_{\rm F}^2)/k_{\rm F}^3\cdot\hat{\bm \sigma }\hat{\sigma }_y,
\end{align}
with the Pauli matrix $\hat{\bm \sigma }$.
When we consider the singly quantized vortex state described by $\Delta({\bm r})=\Delta(\rho)e^{i\phi }$, the spin-degenerate eigenvalue of the CdGM modes is given by~\cite{supplement}
\begin{align}
E_{\nu }=-l\omega_q,
\end{align}
where
\begin{align}
\omega_q\equiv|5\cos^2\alpha -1|\frac{\int_0^{\infty }\frac{|\Delta(\rho')|}{k_{\rm F}\rho'}e^{-2\chi_q(\rho')}d\rho'}{\int_0^{\infty }e^{-2\chi_q(\rho')}d\rho'},
\end{align}
with
\begin{align}
\chi_q(\rho)=\frac{|5\cos^2\alpha -1|}{\hbar v_{\rm F}}\int_0^{\rho }|\Delta(\rho')|d\rho',
\end{align}
by using the Fermi velocity $v_{\rm F}$.
The quantum number $\nu=(l,q)$ consists of the angular momentum $l$ and the wave number along the vortex line $q\equiv k_{\rm F}\cos\alpha$.
The eigenfunction of the CdGM modes for the up- (down-)spin state is given by~\cite{supplement}
\begin{align}
\vec{u}_{\nu }^{\uparrow(\downarrow)}({\bm r})=\widehat{U}_l(\phi)\vec{u}_{\nu }^{\uparrow(\downarrow)}(\rho)e^{iqz}/\sqrt{2\pi },
\label{eq:wavefn}
\end{align}
where $\widehat{U}_l(\phi)\equiv{\rm diag}(e^{i(l+1)\phi },e^{il\phi },e^{i(l-1)\phi },e^{il\phi })/\sqrt{2\pi }$ and
\begin{equation}
\begin{split}
\vec{u}_{\nu }^{\uparrow }(\rho)\equiv&
\mathcal{N}_{\nu }^{\uparrow }
\begin{pmatrix}
J_{l+1}(k_q\rho) \\
0 \\
s_qJ_{l-1}(k_q\rho) \\
0 \\
\end{pmatrix}
e^{-\chi_q(\rho)},\\
\vec{u}_{\nu }^{\downarrow }(\rho)\equiv&
\mathcal{N}_{\nu }^{\downarrow }
\begin{pmatrix}
0 \\
J_{l}(k_q\rho) \\
0 \\
-s_qJ_{l}(k_q\rho) \\
\end{pmatrix}
e^{-\chi_q(\rho)},
\label{eq:wavefn2}
\end{split}
\end{equation}
with the Bessel function $J_l$, $k_q\equiv k_{\rm F}\sin\alpha$, $s_q={\rm sgn}(5\cos^2\alpha-1)$, and the normalization factor $\mathcal{N}_{\nu }^{\uparrow(\downarrow)}$ for the up- (down-)spin state.
Here, we show the eigenvalue $E_{\nu}$ calculated for $\Delta(\rho)=\Delta_0\tanh(\rho/\xi)$ with $k_{\rm F}\xi=5$ in Fig.~\ref{fig1}, where the coherence length is defined by $\xi\equiv\hbar v_{\rm F}/\Delta_0$.

Next, we consider impurity effects on the CdGM modes.
The Dyson equation obeyed by the Matsubara Green's function with impurity self-energy $\widehat{\Sigma }_{\rm imp}({\bm r}_1,{\bm r}_2,\omega_n)$ is described by
\begin{multline}
\widehat{G}({\bm r},{\bm r}',\omega_n)=\widehat{G}^{(0)}({\bm r},{\bm r}',\omega_n)\\
+\int d{\bm r}_1\int d{\bm r}_2 \widehat{G}^{(0)}({\bm r},{\bm r}_1,\omega_n)\widehat{\Sigma }_{\rm imp}({\bm r}_1,{\bm r}_2,\omega_n)\widehat{G}({\bm r}_2,{\bm r}',\omega_n).
\label{eq:Dyson}
\end{multline}
The impurity self-energy is given by
\begin{align}
\widehat{\Sigma }_{\rm imp}({\bm r}_1,{\bm r}_2,\omega_n)=&\frac{\Gamma_{\rm n}}{\pi N_{\rm F}}F({\bm r}_1-{\bm r}_2)\widehat{G}({\bm r}_1,{\bm r}_2,\omega_n)\nn\\
\equiv&\gamma F({\bm r}_1-{\bm r}_2)\widehat{G}({\bm r}_1,{\bm r}_2,\omega_n),
\end{align}
where $N_{\rm F}$ is the density of states (DOS) per spin in the normal state at the Fermi energy, $\Gamma_{\rm n}$ is the impurity scattering rate in the normal state, and $F({\bm r}_1-{\bm r}_2)$ describes the spatial dependence of the squared impurity potential.
Here, $F({\bm r}_1-{\bm r}_2)=\delta({\bm r}_1-{\bm r}_2)$ gives the impurity self-energy by the self-consistent Born approximation~\cite{kopnin:1976,masaki:2016,masaki:2015}.
Instead, we introduce the spatial dependence of the impurity potential in the $z$-direction as $F({\bm r}_1-{\bm r}_2)=\delta({\bm \rho }_1-{\bm \rho }_2)f(z_1-z_2)$, where ${\bm \rho}$ indicates the two-dimensional coordinates perpendicular to the vortex line.
The Green's function without impurities is derived from the BdG wave function in Eq.~\eqref{eq:BdGwave} as
\begin{align}
\widehat{G}^{(0)}({\bm r},{\bm r}',\omega_n)=\sum_{\nu }\frac{\widehat{\tau }_z\vec{u}_{\nu }({\bm r})\vec{u}_{\nu }^{\dagger }({\bm r}')}{E_{\nu }-i\omega_n},
\end{align}
where $\widehat{\bm\tau }$ is the Pauli matrix in the Nambu space.
By deriving $\widehat{G}({\bm r},{\bm r},\omega_n)$ from Eq.~\eqref{eq:Dyson}, we can obtain the DOS as
\begin{align}
N(\omega)=&\int d{\bm r}{\rm Im}\left[\frac{1}{\pi }{\rm Tr}\widehat{\tau }_z\left.\widehat{G}({\bm r},{\bm r},\omega_n)\right|_{i\omega_n\to\omega+i0^+}\right]\nn\\
=&-\frac{1}{\pi }\sum_{\nu }{\rm Im}\left[\frac{1}{\omega -E_{\nu }+\sigma_{\nu }(\omega)+i0^+}\right]\nn\\
\equiv&\sum_{\nu }N_{\nu }(\omega).
\end{align}
Here, we consider the DOS of the CdGM modes $N_{l,q}(\omega)=N_{l,q}^{\uparrow }(\omega)+N_{l,q}^{\downarrow }(\omega)$ by the approximation of neglecting the contributions from the continuum state to the impurity self-energy of the CdGM modes.
Within the approximation, the relaxation time of the CdGM modes depends on the modified self-energy $\sigma_{\nu }^{\uparrow(\downarrow)}(\omega)$ for the up- (down-)spin mode:~\cite{masaki:2016}
\begin{multline}
\sigma_{\nu }^{\uparrow(\downarrow)}(\omega)=-\frac{\gamma }{(2\pi)^2}\sum_{l'}\int dq'\tilde{f}(q-q')\\
\times\frac{M_{\nu,\nu'}^{\uparrow(\downarrow)}}{\omega -E_{\nu'}+\sigma_{\nu'}^{\uparrow(\downarrow)}(\omega)+i0^+},
\label{eq:self-energy}
\end{multline}
where $\tilde{f}(q)$ is the Fourier transform of $f(z)$.
The modified self-energy reflects the wave function of the CdGM modes in Eq.~\eqref{eq:wavefn2} through the coherence factor
\begin{align}
M_{\nu,\nu'}^{\uparrow(\downarrow)}=&\int\rho d\rho\left|\vec{u}_{\nu }^{\uparrow(\downarrow)\dagger }(\rho)\widehat{\tau }_z\vec{u}_{\nu'}^{\uparrow(\downarrow)}(\rho)\right|^2\nn\\
=&\int\rho d\rho\left|m_{\nu,\nu'}^{\uparrow(\downarrow)}(\rho)\right|^2e^{-2\chi_q(\rho)-2\chi_{q'}(\rho)},
\end{align}
where
\begin{equation}
\begin{split}
&m_{\nu,\nu'}^{\uparrow }(\rho)=\mathcal{N}_{\nu }^{\uparrow }\mathcal{N}_{\nu'}^{\uparrow }\\
&\times[J_{l+1}(k_q\rho)J_{l'+1}(k_{q'}\rho)-s_qs_{q'}J_{l-1}(k_q\rho)J_{l'-1}(k_{q'}\rho)],\\
&m_{\nu,\nu'}^{\downarrow }(\rho)=\mathcal{N}_{\nu }^{\downarrow }\mathcal{N}_{\nu'}^{\downarrow }\\
&\times[J_{-l}(k_q\rho)J_{-l'}(k_{q'}\rho)-s_qs_{q'}J_{-l}(k_q\rho)J_{-l'}(k_{q'}\rho)].
\end{split}
\end{equation}
%

Altogether, the coherence factor dominates the relaxation time of the CdGM modes.
If $M_{\nu,\nu'}=0$, the $\nu'$th CdGM mode does not contribute to the modified self-energy on the $\nu$th CdGM mode.
When $\nu$th and $\nu'$th CdGM modes have momentum inside or outside the horizontal line nodes, namely $s_qs_{q'}=1$, $m_{\nu,\nu'}^{\downarrow }(\rho)=0$ irrespective of $l$ and $l'$.
Thus, impurity scattering with the transition between these CdGM modes in the down-spin state does not occur, which is similar to the impurity effects on the CdGM modes in an antiparallel vortex for the chiral $p$-wave state~\cite{kato:2002,masaki:2016}.
Here, ``antiparallel" indicates that the vorticity is antiparallel to the orbital chirality.

On the other hand, $m_{\nu,\nu'}^{\uparrow }(\rho)\ne 0$ even for $s_qs_{q'}=1$ except when $l'=-l$.
If the impurity scattering rate of the CdGM modes $\Gamma$ is smaller than the level spacing of the eigenvalue, the CdGM modes are dominantly transferred between states with the same angular momentum.
For $s_qs_{q'}=1$, since $m_{(0,q),(0,q')}^{\uparrow }=0$, the zero-energy CdGM modes in the up-spin state are hardly scattered, which is similar to the impurity effects on the CdGM modes in a parallel vortex for the chiral $p$-wave state~\cite{masaki:2016}.
The different contributions from the up- and down-spin states to the coherence factor are similar to the case of an $s$-wave topological superconductor by Rashba-type spin orbit coupling~\cite{masaki:2015}.

Thus, impurity scattering with the transition between the zero-energy CdGM modes with $s_qs_{q'}=1$ hardly occurs;
however, the scattering between the CdGM modes with $s_qs_{q'}=-1$ generally occurs, regardless of whether the excitation energy is zero or not, because $M_{\nu,\nu'}\ne 0$.


\begin{figure}
\begin{center}
\includegraphics[width=8.5cm]{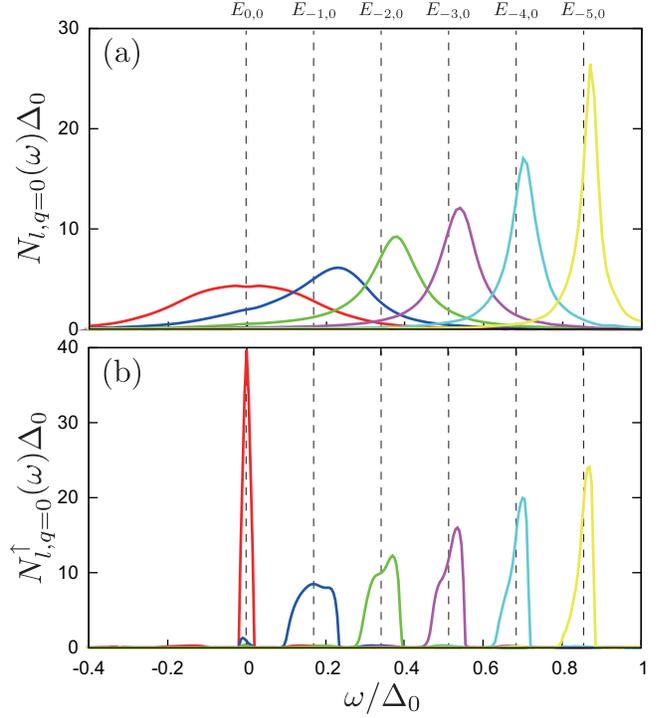}
\end{center}
\caption{\label{fig2}(Color online)
DOS of CdGM modes $N_{l,q=0}$ for $-5\le l \le 0$ derived using the modified self-energy in Eq.~\eqref{eq:self-energy} with $\tilde{f}(q-q')=1$ (a) and $\tilde{f}(q-q')=\theta(k_{\rm F}/10-|q-q'|)$ (b).
}
\end{figure}

In Fig.~\ref{fig2}, we show the DOS of each CdGM mode $N_{l,q=0}$ for $-5\le l\le 0$, where the impurity scattering rate $\Gamma$ of the CdGM mode is reflected by the width of the DOS spectrum.
In the calculation of the modified self-energy in Eq.~\eqref{eq:self-energy}, we sum the CdGM modes with $|l'|\le 10$ and set the impurity scattering rate in the normal state to $\Gamma_{\rm n}=0.05\pi\Delta_0$.
We again take $\Delta(\rho)=\Delta_0\tanh(\rho/\xi)$ with $k_{\rm F}\xi=5$, which fixes $\chi_q(\rho)$ in the coherence factor and the eigenvalue of the CdGM modes $E_{\nu'}$ as shown in Fig.~\ref{fig1}.
The difference between Figs.~\ref{fig2}(a) and \ref{fig2}(b) is the spatial dependence of the impurity potential in the $z$-direction.

In Fig.~\ref{fig2}(a), we put $\tilde{f}(q-q')=1$ in Eq.~\eqref{eq:self-energy}, which implies $f(z)=\delta(z)$.
Since the modified self-energy is given by the integral over all $|q'|<k_{\rm F}$, there are many pairs of zero-energy CdGM modes with $s_qs_{q'}=-1$ across the line nodes; thus, the many relaxation processes give a large $\Gamma$ for the zero-energy modes.
For $N_{l,q=0}$ with $l<0$, the lower-energy mode has the broader spectrum, which is similar to the CdGM modes in a vortex for the $s$-wave state and in a parallel vortex for the chiral $p$-wave state~\cite{kato:2002,masaki:2016}.

Next, we consider the step function $\tilde{f}(q-q')=\theta(k-|q-q'|)$, which is the Fourier transform of $f(z)=\sin(kz)/(\pi z)$.
In the calculation of the DOS of the CdGM modes in Fig.~\ref{fig2}(b), we take $k=k_{\rm F}/10$.
In this situation, the down-spin CdGM modes have the DOS $N_{l,q=0}^{\downarrow }(\omega)=\delta(\omega -E_{l,0})$ owing to the coherence factor $M_{(l,0),(l',q')}^{\downarrow }=0$ within $|q'|<k_{\rm F}/\sqrt{5}$.
In Fig.~\ref{fig2}(b), we show the DOS $N_{l,q=0}^{\uparrow }$ for the up-spin CdGM modes, in which only the zero-energy mode has a sharp DOS, which is due to the fact that $M_{(0,0),(0,q')}^{\uparrow }=0$.
Except for the zero-energy CdGM mode, the lower-energy mode has the broader spectrum with an asymmetric tail toward the low energy, where the asymmetric spectrum reflects the convex dispersion of the CdGM modes for $l<0$ around $q=0$.

The robust zero-energy modes demonstrated in Fig.~\ref{fig2}(b) can be understood from the one-dimensional winding number, which is a topological number defined on a closed path with particular values of momentum.
For the calculation of the winding number, we consider the semiclassical BdG Hamiltonian parametrized by the angle $\phi$ on a circle surrounding the vortex as
\begin{align}
\widehat{\mathcal{H}}({\bm k},\phi)=
\begin{pmatrix}
\hat{\epsilon }(\bm k) & \hat{\Delta }({\bm k},\phi) \\
\hat{\Delta }^{\dagger }({\bm k},\phi) & -\hat{\epsilon }^{\rm T}(-\bm k)
\end{pmatrix}.
\end{align}
Here, $\hat{\epsilon }(\bm k)$ is the Hamiltonian in the normal state of UPt$_3$, which has the D$_{\rm 6h}$ hexagonal symmetry.
For the $E_{1u}$ planar state with a vortex, whose pair potential is described by $\hat{\Delta }({\bm k},\phi)=\Delta_0e^{i\phi }(-k_y\hat{\sigma }_z+ik_x\hat{\sigma }_0)(5k_z^2-k_{\rm F}^2)/k_{\rm F}^3$, the semiclassical BdG Hamiltonian has magnetic reflection symmetry on the $xz$-plane~\cite{tsutsumi:2013}.
From the combination of the magnetic reflection symmetry and particle-hole symmetry, we obtain chiral symmetry on $k_y=0$ and $\phi=0$ or $\pi$ as $\{\widehat{\Gamma },\widehat{\mathcal{H}}_{\phi=0,\pi}\}=0$, where $\widehat{\Gamma }\equiv\widehat{\tau }_x$ and $\widehat{\mathcal{H}}_{\phi=0,\pi}\equiv\widehat{\mathcal{H}}(k_x,k_y=0,k_z,\phi=0,\pi)$.
The chiral symmetry enables us to define the one-dimensional winding number as~\cite{sato:2009c,sato:2011}
\begin{align}
w_{\phi=0,\pi }(k_z)=-\frac{1}{4\pi i}\int_{\rm BZ}dk_x{\rm Tr}\left[\widehat{\Gamma }\widehat{\mathcal{H}}_{\phi=0,\pi}^{-1}\partial_{k_x}\widehat{\mathcal{H}}_{\phi=0,\pi}\right],
\end{align}
where the line integral should be performed over the Brillouin zone.
Since the winding number depends on the sign of the gap function~\cite{sato:2011}, we evaluate $w_0=-w_{\pi }=-2$ for $|k_z|<k_{\rm F}/\sqrt{5}$, $w_0=-w_{\pi }=2$ for $k_{\rm F}/\sqrt{5}<|k_z|<k_{\rm F}$, and $w_0=w_{\pi }=0$ for $|k_z|>k_{\rm F}$~\cite{tsutsumi:2013}.
The difference in the winding number $(w_0-w_{\pi })/2=N$ guarantees the existence of at least $|N|$ zero-energy modes~\cite{shiozaki:2014,tsutsumi:2015}.
The sign of the winding number is related to the eigenvalue of $\widehat{\Gamma }$ for the zero-energy mode, namely its chirality.
Impurity effects of the short-range potential in the $k_z$-space do not affect the zero-energy CdGM modes around $q=0$ with the same chirality.
This is because perturbations couple the zero-energy modes with opposite chirality~\cite{sato:2011}.

Since we have already demonstrated that the topological argument of nodal superconductors is effective even when momentum transfer is caused by impurity scattering, we can discuss the robustness of the CdGM modes against impurities for the other possible superconducting states of UPt$_3$ without complicated calculations.
The $E_{1u}$ and $E_{2u}$ chiral states have the gap functions $\hat{\Delta }({\bm k},\phi)=\Delta_0e^{i\phi }(k_x+ik_y)(5k_z^2-k_{\rm F}^2)/k_{\rm F}^3\hat{\sigma }_x$ and $\hat{\Delta }({\bm k},\phi)=\Delta_0e^{i\phi }(k_x+ik_y)^2k_z/k_{\rm F}^3\hat{\sigma }_x$, respectively.
Both the gap functions give the winding number $w_0=w_{\pi }=0$ for any $k_z$~\cite{kobayashi:2015} owing to the difference in the spin state from the $E_{1u}$ planar state.
The trivial topological number reflects the absence of zero-energy CdGM modes.
For the $E_{1u}$ chiral state, the lowest CdGM modes indeed have a gap owing to the magnetic field in the $z$-direction.
Since the magnetic field deforms the wave function of the CdGM modes with the energy shift, $M_{(0,q),(0,q)}\ne 0$.
Therefore, the lowest CdGM modes have a broadened DOS even by columnar defects without momentum transfer on $k_z$.

In summary, we have studied impurity effects on CdGM modes in a vortex for nodal topological superconductors by focusing on the $E_{1u}$ planar state, which is a possible superconducting state in UPt$_3$.
We have demonstrated that the coherence factor of the zero-energy CdGM modes vanishes for quasiparticle pairs with momenta on the same side of a horizontal line node.
The coherence factor reflects the one-dimensional winding number defined on a particular momentum space, which does not change unless crossing the horizontal line node.
Thus, the classification of nodal topological superconductors by the topological number on a particular momentum space is effective even when there is momentum transfer.
The characteristic coherence effects of nodal topological superconductors shown in Fig.~\ref{fig2} can be confirmed by comparing flux flow conductivities, which depend on the relaxation time $\tau=\hbar/\Gamma$ of the CdGM modes~\cite{kopnin:book}, before and after spreading artificial columnar defects~\cite{civale:1991,toulemonde:1994}.
We may directly observe the $q$-dependent coherence factor of the zero-energy CdGM modes by inelastic neutron scattering experiments or quasiparticle interference imaging.

\begin{acknowledgments}
We thank T.~Hanaguri, S.~Hoshino, Y.~Masaki, and S.~Kobayashi for helpful discussion.
Y.T.~acknowledges financial support from the Japan Society for the Promotion of Science (JSPS).
This work was supported by JSPS KAKENHI Grant Numbers 15K17715 and 15J05698.

\end{acknowledgments}


\clearpage
\onecolumn

\renewcommand{\thefigure}{S\arabic{figure}} 

\renewcommand{\thetable}{S\arabic{table}} 

\renewcommand{\thesection}{S\arabic{section}.}

\renewcommand{\theequation}{S.\arabic{equation}}

\setcounter{figure}{0}
\setcounter{table}{0}
\setcounter{equation}{0}

\begin{flushleft} 
{\Large {\bf Supplementary Material}}
\end{flushleft}

\begin{flushleft} 
{\bf Caroli-de Gennes-Matricon modes for the $E_{1u}$ planar state}
\end{flushleft} 

In the Supplemental Material, we derive the Caroli-de Gennes-Matricon (CdGM) modes~\cite{scaroli:1964} in a vortex for the $E_{1u}$ planar state~\cite{smachida:2012,stsutsumi:2012b}, which is a strong candidate for the superconducting state of UPt$_3$.
We start out the derivation from the Bogoliubov-de Gennes (BdG) equation for the inhomogeneous order parameter:
\begin{align}
\int d{\bm r}_2
\begin{pmatrix}
\hat{\epsilon }(\bi{r}_1,\bi{r}_2) & \hat{\Delta }(\bi{r}_1,\bi{r}_2) \\
\hat{\Delta }^{\dagger }(\bi{r}_2,\bi{r}_1) & -\hat{\epsilon }^{\rm T}(\bi{r}_2,\bi{r}_1)
\end{pmatrix}
\vec{u}_{\nu }(\bi{r}_2)
=E_{\nu }\vec{u}_{\nu }(\bi{r}_1).
\label{eq:BdG}
\end{align}
The normal state Hamiltonian omitting the vector potential is given by
\begin{align}
\hat{\epsilon }(\bi{r}_1,\bi{r}_2)=\delta(\bi{r}_1-\bi{r}_2)\frac{\hbar^2}{2m}\left(-\partial_{\rho}^2-\frac{1}{\rho}\partial_{\rho}-\frac{1}{\rho^2}\partial_{\phi}^2-\partial_z^2-k_{\rm F}^2\right)\hat{\sigma }_0,
\end{align}
where $m$ is the particle mass, $k_{\rm F}$ is the Fermi wave number, $\hat{\sigma }_0$ is the unit matrix in the spin space, and $(\partial_{\rho },\partial_{\phi },\partial_z)$ are differential operators in cylindrical coordinates.
The pair potential is
\begin{align}
\hat{\Delta}(\bi{r}_1,\bi{r}_2)=\int\frac{d{\bm k}}{(2\pi)^3}\hat{\Delta }({\bm r},{\bm k})e^{i{\bm k}\cdot{\bm r}'},
\label{eq:pairpot}
\end{align}
with ${\bm r}=({\bm r}_1+{\bm r}_2)/2$ and ${\bm r}'={\bm r}_1-{\bm r}_2$,
and the wave function is
\begin{align}
\vec{u}_{\nu }({\bm r})=
\begin{pmatrix}
u_{\nu }^{\uparrow }({\bm r}) \\
u_{\nu }^{\downarrow }({\bm r}) \\
v_{\nu }^{\uparrow }({\bm r}) \\
v_{\nu }^{\downarrow }({\bm r})
\end{pmatrix}.
\end{align}

The gap function for the $E_{1u}$ planar state is described by
\begin{align}
\hat{\Delta }({\bm r},{\bm k})=i\hat{\bm\sigma }\hat{\sigma }_y\cdot\frac{\Delta({\bm r})}{k_{\rm F}^3}({\bm x}k_y+{\bm y}k_x)(5k_z^2-k_{\rm F}^2),
\label{eq:gapfn}
\end{align}
with the Pauli matrix $\hat{\bm \sigma }$.
By substituting Eq.~\eqref{eq:gapfn} into Eq.~\eqref{eq:pairpot} and following the procedure in Ref.~\cite{smatsumoto:2001}, the pair potential is rewritten as
\begin{align}
\hat{\Delta}(\bi{r}_1,\bi{r}_2)=-\frac{i}{k_{\rm F}}\Delta({\bm r})\hat{\mathcal{D}}({\bm r}_2)\delta({\bm r}_1-{\bm r}_2),
\end{align}
where
\begin{align}
\hat{\mathcal{D}}({\bm r})\equiv i\hat{\bm \sigma }\hat{\sigma }_y\cdot({\bm x}\partial_y+{\bm y}\partial_x)\frac{5\partial_z^2+k_{\rm F}^2}{k_{\rm F}^2}.
\end{align}
When we consider a vortex along the $z$-direction, the order parameter $\Delta({\bm r})$ is uniform to the $z$-direction.
By the integral in Eq.~\eqref{eq:BdG}, the BdG equation is reduced to
\begin{align}
\begin{pmatrix}
\hat{\epsilon }(\bi{r}) & \frac{i}{k_{\rm F}}\left[\Delta(\bi{r})\hat{\mathcal{D}}({\bm r})+\frac{1}{2}\hat{\mathcal{D}}({\bm r})\Delta(\bi{r})\right] \\
\frac{i}{k_{\rm F}}\left[\Delta^*(\bi{r})\hat{\mathcal{D}}^{\dagger }({\bm r})+\frac{1}{2}\hat{\mathcal{D}}^{\dagger }({\bm r})\Delta^*(\bi{r})\right] & -\hat{\epsilon }^{\rm T}(\bi{r})
\end{pmatrix}
\vec{u}_{\nu }(\bi{r})
=E_{\nu }\vec{u}_{\nu }(\bi{r}),
\label{eq:BdG2}
\end{align}
where
\begin{align}
\hat{\epsilon }(\bi{r})\equiv\epsilon({\bm r})\hat{\sigma }_0\equiv\frac{\hbar^2}{2m}\left(-\partial_{\rho}^2-\frac{1}{\rho}\partial_{\rho}-\frac{1}{\rho^2}\partial_{\phi}^2-\partial_z^2-k_{\rm F}^2\right)\hat{\sigma }_0.
\end{align}
Here, the BdG equation can be separated into the up-spin state and the down-spin state, where ${\bm u}_{\nu }^{\uparrow }\equiv(u_{\nu }^{\uparrow },v_{\nu }^{\uparrow })^{\rm T}$ under $\Delta({\bm r})$ is equivalent to $\left({\bm u}_{\nu }^{\downarrow }\right)^*\equiv(u_{\nu }^{\downarrow },v_{\nu }^{\downarrow })^{\dagger }$ under $\Delta^*({\bm r})$.

When we consider the axisymmetric vortex state described by $\Delta({\bm r})=\Delta(\rho)\exp(i\kappa\phi )$ with a vorticity $\kappa\in\mathbb{Z}$, we can derive the BdG wave function as
\begin{align}
\vec{u}_{\nu }({\bm r})=\frac{1}{2\pi }
\begin{pmatrix}
u_{\nu }^{\uparrow }(\rho) e^{i\frac{\kappa +1}{2}\phi } \\
u_{\nu }^{\downarrow }(\rho) e^{i\frac{\kappa -1}{2}\phi } \\
v_{\nu }^{\uparrow }(\rho) e^{-i\frac{\kappa +1}{2}\phi } \\
v_{\nu }^{\downarrow }(\rho) e^{-i\frac{\kappa -1}{2}\phi }
\end{pmatrix}
e^{il\phi }e^{iqz},
\label{eq:eigenfunction}
\end{align}
with $\nu=(n,l,q)$ consisting of $n\in\mathbb{Z}$ related to the phase of the wave function, integer $l$ for odd $\kappa$ or half-integer $l$ for even $\kappa$, and $q=k_{\rm F}\cos\alpha$.
The BdG equation for the up-spin state is described by
\begin{multline}
\begin{pmatrix}
\epsilon({\bm r}) & \frac{i}{k_{\rm F}}\left[\Delta({\bm r})\mathcal{D}({\bm r})+\frac{1}{2}\mathcal{D}({\bm r})\Delta({\bm r})\right] \\
\frac{i}{k_{\rm F}}\left[\Delta^*({\bm r})\mathcal{D}^*({\bm r})+\frac{1}{2}\mathcal{D}^*({\bm r})\Delta^*({\bm r})\right] & -\epsilon({\bm r})
\end{pmatrix}
\begin{pmatrix}
u_{\nu }^{\uparrow}(\rho)e^{i\frac{\kappa +1}{2}\phi } \\ v_{\nu }^{\uparrow}(\rho)e^{-i\frac{\kappa +1}{2}\phi }
\end{pmatrix}
e^{il\phi }e^{iqz}\\
=E_{\nu }
\begin{pmatrix}
u_{\nu }^{\uparrow}(\rho)e^{i\frac{\kappa +1}{2}\phi } \\ v_{\nu }^{\uparrow}(\rho)e^{-i\frac{\kappa +1}{2}\phi }
\end{pmatrix}
e^{il\phi }e^{iqz},
\label{eq:BdGup}
\end{multline}
where
\begin{align}
\mathcal{D}({\bm r})\equiv ie^{i\phi }\left(\partial_{\rho }+i\frac{1}{\rho }\partial_{\phi }\right)\frac{5\partial_z^2+k_{\rm F}^2}{k_{\rm F}^2}.
\end{align}
By the symmetry between the up-spin state and the down-spin state, eigenvalue and eigenfunction of the down-spin state are given by $E_{n,-l,-q}$ and $\left({\bm u}_{n,-l,-q}^{\uparrow }\right)^*$, respectively, under the order parameter with the opposite vorticity $-\kappa$.
From now on, we only consider the up-spin wave function ${\bm u}_{\nu }(\rho)\equiv{\bm u}_{\nu }^{\uparrow }(\rho)$ which has similar features to the chiral $p$-wave state~\cite{skopnin:1991,sstone:2006,smizushima:2010} except for the $q$-dependence.
The following derivation of the CdGM modes obeys Ref.~\cite{smizushima:2010}.
By taking real $\Delta(\rho)\ge 0$, Eq.~\eqref{eq:BdGup} is rewritten as
\begin{multline}
\left[\frac{\epsilon_{\rm F}}{k_{\rm F}^2}\mathcal{L}_m(\alpha)\hat{\tau }_0
-\frac{1}{k_{\rm F}}\left\{\left(\Delta(\rho)\frac{d}{d\rho }+\frac{\Delta(\rho)}{2\rho }\right)(5\cos^2\alpha -1)+\frac{5}{2}\frac{\kappa }{\rho }\Delta(\rho)\cos^2\alpha -\frac{1}{2}\Delta'(\rho)\right\}\hat{\tau }_x
+E_{\nu }\hat{\tau }_z\right]{\bm u}_{\nu }(\rho)\\
=\left[\epsilon_{\rm F}\frac{(\kappa+1)l}{(k_{\rm F}\rho)^2}\hat{\tau }_z-i\Delta(\rho)\frac{l}{k_{\rm F}\rho }(5\cos^2\alpha -1)\hat{\tau }_y\right]{\bm u}_{\nu }(\rho),
\label{eq:BdGup2}
\end{multline}
where $\epsilon_{\rm F}$ is the Fermi energy, $\hat{\tau }_0$ and $\hat{\bm\tau }$ are the unit matrix and the Pauli matrix, respectively, in the Nambu space, and
\begin{align}
\mathcal{L}_m(\alpha)\equiv\frac{d^2}{d^2\rho }+\frac{1}{\rho }\frac{d}{d\rho }-\frac{m^2}{\rho^2}+k_{\rm F}^2\sin^2\alpha,
\end{align}
with $m\equiv\sqrt{l^2+\frac{(\kappa+1)^2}{4}}$.

Here, we introduce a radius $\rho_{\rm c}$ that $\Delta(\rho)=0$ for $\rho<\rho_{\rm c}$, where $|l|\ll k_{\rm F}\rho_{\rm c}\ll k_{\rm F}\xi$ on the coherence length $\xi\equiv\hbar v_{\rm F}/\Delta(\infty)$ defined by using the Fermi velocity $v_{\rm F}$.
The solution of the wave function in Eq.~\eqref{eq:BdGup2} is obtained in the range of $\rho<\rho_{\rm c}$ as
\begin{align}
{\bm u}_{\nu }(\rho)=
\begin{pmatrix}
\mathcal{N}_{\nu }^uJ_{l+\frac{\kappa +1}{2}}(k_{\nu }^+\rho) \\ \mathcal{N}_{\nu }^vJ_{l-\frac{\kappa +1}{2}}(k_{\nu }^-\rho)
\end{pmatrix},
\label{eq:ur<rc}
\end{align}
where $\mathcal{N}_{\nu }^u$ and $\mathcal{N}_{\nu }^v$ are the normalization factors and we define
\begin{align}
k_{\nu }^{\pm }\equiv k_{\rm F}\sin\alpha\pm\frac{E_{\nu }}{\hbar v_{\rm F}\sin\alpha }.
\end{align}

For $\rho>\rho_{\rm c}$, the wave function in Eq.~\eqref{eq:BdGup2} is assumed to consist of the Hankel function $H_m^{(1)}$ and the slow functions varying over the order of $\xi$, ${\bm \varphi }_{\nu }(\rho)$, as
\begin{align}
{\bm u}_{\nu }(\rho)=H_m^{(1)}(k_{\rm F}\sin\alpha\rho){\bm \varphi }_{\nu }(\rho)+{\rm c.c.}
\label{eq:ur>rc}
\end{align}
By using the ordinary condition $k_{\rm F}\xi\gg 1$, Eq.~\eqref{eq:BdGup2} is reduced to
\begin{multline}
\left[\frac{d}{d\rho }\hat{\tau }_0-\frac{\Delta(\rho)}{\hbar v_{\rm F}}\left\{(5\cos^2\alpha -1)-i\frac{5\kappa\cos^2\alpha }{2k_{\rm F}\sin\alpha\rho }\right\}\hat{\tau }_x\right]{\bm \varphi }_{\nu }(\rho)\\
=\left\{i\left[\frac{E_{\nu }}{\hbar v_{\rm F}\sin\alpha }-\frac{(\kappa+1)l}{2k_{\rm F}\sin\alpha\rho^2}\right]\hat{\tau }_z-\frac{\Delta(\rho)}{\hbar v_{\rm F}}\frac{l}{k_{\rm F}\sin\alpha\rho }(5\cos^2\alpha -1)\hat{\tau }_y\right\}{\bm \varphi }_{\nu }(\rho).
\label{eq:varphi}
\end{multline}
By the assumptions $|E_{\nu }|\ll\Delta(\infty)$ and $k_{\rm F}\rho_{\rm c}\gg |l|$, the right-hand-side of Eq.~\eqref{eq:varphi} can be regarded as a small perturbation.
The solution within the first order on $|\psi_{\nu }|\ll 1$ is given by
\begin{align}
{\bm \varphi }_{\nu }(\rho)=A_{\nu }e^{-\chi_q(\rho)+i\kappa\tilde{\chi }_q(\rho)}
\begin{pmatrix}
1+i\psi_{\nu }(\rho) \\ -s_q[1-i\psi_{\nu }(\rho)]
\end{pmatrix}\approx A_{\nu }e^{-\chi_q(\rho)+i\kappa\tilde{\chi }_q(\rho)}
\begin{pmatrix}
e^{i\psi_{\nu }(\rho)} \\ -s_q e^{-i\psi_{\nu }(\rho)}
\end{pmatrix},
\label{eq:varphisolution}
\end{align}
where $s_q\equiv{\rm sgn}(5\cos^2\alpha -1)$,
\begin{align}
\chi_q(\rho)\equiv\frac{|5\cos^2\alpha -1|}{\hbar v_{\rm F}}\int_0^{\rho }\Delta(\rho')d\rho',\
\tilde{\chi }_q(\rho)\equiv s_q\frac{5\cos^2\alpha }{2\hbar v_{\rm F}\sin\alpha }\int_0^{\rho }\frac{\Delta(\rho')}{k_{\rm F}\rho'}d\rho',
\end{align}
and
\begin{align}
\psi_{\nu }(\rho)\equiv -\int_{\rho }^{\infty }\left[\frac{E_{\nu }}{\hbar v_{\rm F}\sin\alpha }-\frac{(\kappa+1)l}{2k_{\rm F}\sin\alpha\rho'^2}-\frac{\Delta(\rho')}{\hbar v_{\rm F}}\frac{l}{k_{\rm F}\sin\alpha\rho' }|5\cos^2\alpha -1|\right]e^{-2[\chi_q(\rho')-\chi_q(\rho)]}d\rho'.
\label{eq:psi}
\end{align}

In order to obtain the solution of Eq.~\eqref{eq:BdGup2}, the wave functions in Eq.~\eqref{eq:ur<rc} for $\rho<\rho_{\rm c}$ and in Eq.~\eqref{eq:ur>rc} for $\rho>\rho_{\rm c}$ are matched at $\rho=\rho_{\rm c}$.
By using the asymptotic forms of $J_m(x)$ and $H_m^{(1)}(x)$ in $x\gg|m|$, the wave function in Eq.~\eqref{eq:ur<rc} becomes
\begin{align}
{\bm u}_{\nu }(\rho_{\rm c})=\sqrt{\frac{2}{\pi k_{\rm F}\sin\alpha\rho_{\rm c}}}
\begin{pmatrix}
\mathcal{N}_{\nu }^u\cos\left[k_{\nu }^+\rho_{\rm c}+\frac{(l+\frac{\kappa+1}{2})^2-1/4}{2k_{\rm F}\sin\alpha\rho_{\rm c}}-\frac{2(l+\frac{\kappa+1}{2})+1}{4}\pi\right] \\
\mathcal{N}_{\nu }^v\cos\left[k_{\nu }^-\rho_{\rm c}+\frac{(l-\frac{\kappa+1}{2})^2-1/4}{2k_{\rm F}\sin\alpha\rho_{\rm c}}-\frac{2(l-\frac{\kappa+1}{2})+1}{4}\pi\right]
\end{pmatrix}.
\label{eq:r=rc}
\end{align}
at $\rho=\rho_{\rm c}$ and the wave function in Eq.~\eqref{eq:ur>rc} is rewritten as
\begin{align}
{\bm u}_{\nu }(\rho)=\sqrt{\frac{2}{\pi k_{\rm F}\sin\alpha\rho }}e^{-\chi_q(\rho)}
\begin{pmatrix}
A_{\nu }e^{i\eta_{\nu }^+(\rho)}+A_{\nu }^*e^{-i\eta_{\nu }^+(\rho)} \\
-s_q[A_{\nu }e^{i\eta_{\nu }^-(\rho)}+A_{\nu }^*e^{-i\eta_{\nu }^-(\rho)}]
\end{pmatrix},
\label{eq:r>rc}
\end{align}
for $\rho\ge\rho_{\rm c}$, where
\begin{align}
\eta_{\nu }^{\pm }(\rho)\equiv k_{\rm F}\sin\alpha\rho+\frac{m^2-1/4}{2k_{\rm F}\sin\alpha\rho }-\frac{2m+1}{4}\pi+\kappa\tilde{\chi }_q(\rho)\pm\psi_{\nu }(\rho).
\label{eq:eta}
\end{align}
Since $\chi_q(\rho_{\rm c})=\tilde{\chi }_q(\rho_{\rm c})=0$, in order to match two expressions ${\bm u}_{\nu }(\rho_{\rm c})$ in Eq.~\eqref{eq:r=rc} and in Eq.~\eqref{eq:r>rc} at $\rho=\rho_{\rm c}$, $\psi_{\nu }(\rho_{\rm c})$ in Eq.~\eqref{eq:eta} should be satisfied both conditions
\begin{align}
\psi_{\nu }(\rho_{\rm c})=\frac{E_{\nu }}{\hbar v_{\rm F}\sin\alpha }\rho_{\rm c}+\frac{(\kappa+1)l}{2k_{\rm F}\sin\alpha\rho_{\rm c}}+\frac{\pi }{2}\left(m-l-\frac{\kappa+1}{2}\right) -\gamma_{\nu },
\label{eq:psir=rc1}
\end{align}
and
\begin{align}
\psi_{\nu }(\rho_{\rm c})=\frac{E_{\nu }}{\hbar v_{\rm F}\sin\alpha }\rho_{\rm c}+\frac{(\kappa+1)l}{2k_{\rm F}\sin\alpha\rho_{\rm c}}-\frac{\pi }{2}\left(m-l+\frac{\kappa+1}{2}\right) +\gamma_{\nu } +n\pi,
\label{eq:psir=rc2}
\end{align}
with the relation among the coefficients, that is $\mathcal{N}_{\nu }^v=-s_q\mathcal{N}_{\nu }^u$ for even $n$ or $\mathcal{N}_{\nu }^v=s_q\mathcal{N}_{\nu }^u$ for odd $n$ and $A_{\nu }=(\mathcal{N}_{\nu }^u/2)e^{i\gamma_{\nu }}$.
When $\gamma_{\nu }=(m-l-n)\pi/2$, Eqs.~\eqref{eq:psir=rc1} and \eqref{eq:psir=rc2} become the identical expression:
\begin{align}
\psi_{\nu }(\rho_{\rm c})=\frac{E_{\nu }}{\hbar v_{\rm F}\sin\alpha }\rho_{\rm c}+\frac{(\kappa+1)l}{2k_{\rm F}\sin\alpha\rho_{\rm c}}+\frac{\pi }{2}\left(n-\frac{\kappa+1}{2}\right).
\label{eq:psir=rc3}
\end{align}

The general expression of $\psi_{\nu }(\rho)$ in Eq.~\eqref{eq:psi} for $\rho>\rho_{\rm c}$ should correspond to Eq.~\eqref{eq:psir=rc3} at $\rho=\rho_{\rm c}$.
This condition provides the eigenvalue of the BdG equation~\eqref{eq:BdGup} as
\begin{align}
E_{\nu }=-\kappa l\omega_q-\left(n-\frac{\kappa+1}{2}\right)\omega_q',
\end{align}
where
\begin{align}
\omega_q\equiv|5\cos^2\alpha -1|\frac{\int_0^{\infty }\frac{\Delta(\rho')}{k_{\rm F}\rho'}e^{-2\chi_q(\rho')}d\rho'}{\int_0^{\infty }e^{-2\chi_q(\rho')}d\rho'}\sim\frac{\Delta(\infty)^2}{\epsilon_{\rm F}},
\end{align}
and
\begin{align}
\omega_q'\equiv\sin\alpha\frac{\pi\hbar v_{\rm F}}{2\int_0^{\infty }e^{-2\chi_q(\rho')}d\rho'}\sim\Delta(\infty).
\end{align}
Since $\omega_q\ll\omega_q'$, the eigenvalue of the CdGM modes for singly quantized vortex states with $\kappa=\pm 1$ is given by $n=(\kappa+1)/2$~\cite{ssuzuki:2008}.

By using the eigenvalue $E_{\nu }=-\kappa l\omega_q$, the phase $\psi_{\nu }(\rho)$ of the eigenfunction of the CdGM modes for $\kappa=\pm 1$ is given by Eq.~\eqref{eq:psi} as
\begin{align}
\psi_{\nu }(\rho)=\frac{E_{\nu }}{\hbar v_{\rm F}\sin\alpha }\rho+\frac{(\kappa+1)l}{2k_{\rm F}\sin\alpha\rho }+\kappa l\tilde{\psi }_q(\rho),
\end{align}
where
\begin{align}
\tilde{\psi }_q(\rho)\equiv\frac{|5\cos^2\alpha -1|}{\hbar v_{\rm F}\sin\alpha }
\int_0^{\rho }\left[\frac{1}{k_{\rm F}\rho'}-\frac{2\omega_q}{\hbar v_{\rm F}}\rho'\right]\Delta(\rho')e^{-2[\chi_q(\rho')-\chi_q(\rho)]}d\rho'.
\end{align}
From Eq.~\eqref{eq:r>rc}, the eigenfunction for the up-spin state is given by
\begin{align}
{\bm u}_{\nu }^{\uparrow }(\rho)=\mathcal{N}_{\nu }^{\uparrow }
\begin{pmatrix}
J_{l+\frac{\kappa +1}{2}}[k_{\nu }^+\rho+\kappa l\tilde{\psi }_q(\rho)+\kappa\tilde{\chi }_q(\rho)] \\
\kappa s_qJ_{l-\frac{\kappa +1}{2}}[k_{\nu }^-\rho -\kappa l\tilde{\psi }_q(\rho)+\kappa\tilde{\chi }_q(\rho)] \\
\end{pmatrix}e^{-\chi_q(\rho)},
\end{align}
which is matched to Eq.~\eqref{eq:ur<rc} at $\rho=\rho_{\rm c}$ because $\chi_q(\rho_{\rm c})=\tilde{\chi }_q(\rho_{\rm c})=\tilde{\psi }_q(\rho_{\rm c})=0$ due to the assumption $\Delta(\rho<\rho_{\rm c})=0$.
The evaluation of the argument in the Bessel function can be advanced to
\begin{align}
k_{\nu }^{\pm }\rho\pm\kappa l\tilde{\psi }_q(\rho)+\kappa\tilde{\chi }_q(\rho)=&k_{\rm F}\sin\alpha\rho\pm\frac{E_{\nu }}{\hbar v_{\rm F}\sin\alpha }\int_0^{\rho }\left(1-\frac{|5\cos^2\alpha -1|}{\omega_q}\frac{\Delta(\rho')}{k_{\rm F}\rho'}\right)e^{-2[\chi_q(\rho')-\chi_q(\rho)]}d\rho'+\kappa\tilde{\chi }_q(\rho)\nn\\
\equiv&k_{\rm F}\sin\alpha\rho\pm\frac{E_{\nu }}{\hbar v_{\rm F}\sin\alpha }f_q(\rho)+\kappa\tilde{\chi }_q(\rho).
\label{eq:argument}
\end{align}
Since $|f_q(\rho)|/\xi\lesssim 1$, the second term in Eq.~\eqref{eq:argument} can be neglected by $|E_{\nu }|\ll\Delta(\infty)$.
Moreover, the third term in Eq.~\eqref{eq:argument} is much less than the first term for any $\rho$ owing to $k_{\rm F}\xi\gg 1$.
Finally, from the symmetry between ${\bm u}_{\nu}^{\uparrow }$ and ${\bm u}_{\nu}^{\downarrow }$, the BdG wave function in Eq.~\eqref{eq:eigenfunction} is given by
\begin{align}
\begin{pmatrix}
u_{\nu }^{\uparrow }(\rho) \\
v_{\nu }^{\uparrow }(\rho)
\end{pmatrix}
=\mathcal{N}_{\nu }^{\uparrow }
\begin{pmatrix}
J_{l+\frac{\kappa +1}{2}}(k_q\rho) \\
\kappa s_qJ_{l-\frac{\kappa +1}{2}}(k_q\rho)
\end{pmatrix}e^{-\chi_q(\rho)},\
\begin{pmatrix}
u_{\nu }^{\downarrow }(\rho) \\
v_{\nu }^{\downarrow }(\rho)
\end{pmatrix}
=\mathcal{N}_{\nu }^{\downarrow }
\begin{pmatrix}
J_{l+\frac{\kappa -1}{2}}(k_q\rho) \\
-\kappa s_qJ_{l-\frac{\kappa -1}{2}}(k_q\rho)
\end{pmatrix}e^{-\chi_q(\rho)},
\end{align}
with $k_q\equiv k_{\rm F}\sin\alpha$.

\end{document}